**DD-03B: Target-level analysis of ligand–protein dissociation pathways reveals ligand-specific exit pathways weakly coupled to binding affinity**

**Authors:** Maodong Li, Keyu Gao, Yiming Chen, Zhe Wang, Runxiu Wu, Dechin Chen, Zhijun Pan, Yi Isaac Yang*

**Affiliations:** Institute of Systems and Physical Biology, Shenzhen Bay Laboratory, Shenzhen 518107, China.

*e-mail: yangyi@szbl.ac.cn

## Abstract

Protein–ligand interactions are conventionally described by a structure and a binding affinity (Kd). Kinetic simulation datasets have grown rapidly, but most analyses remain structure-centred and RMSD-based, and the kinetic information carried by the trajectories is rarely extracted. Here we analyse 766,550 complete dissociation trajectories for 15,540 protein–ligand complexes grouped into 3,687 UniProt accessions. Each trajectory is reduced to a uniform record of 119 descriptors — ligand and structure size, trajectory statistics, the per-pathway distribution of escape events, and pocket and escape-volume descriptors — so that records from different systems can be compared, joined and searched. This tokenisation shows that the pocket and the ligand act on the same quantity at different levels: the pocket defines the repertoire of exits, and the ligand sets the weights. In open sites the repertoire collapses to one broad mode and affinity tracks ligand size and contact; in confined pockets the exit ensemble splits into several channels whose weights follow ligand size rather than binding affinity. UniProt grouping is a convenient classification above the PDB entry but not the right final unit: two systems illustrate domain separation and degeneracy, and a structural classifier would be preferable. We present the framework as a general analysis paradigm for molecular dynamics trajectory libraries.

Protein–ligand interactions are a central object of biological research, and the established way to study them is structural: solve the structure, locate the pocket, dock the ligand, score the pose, and calibrate the score against measured binding affinities[3–5]. For the thermodynamics of binding this pipeline has been remarkably successful — a structure plus an affinity is often enough to explain why a compound binds and how tightly. It does not describe kinetics. The residence time of a ligand on its target, and the dissociation rate constant koff that sets it, can change the efficacy of a drug at a given dose[1,2], yet neither follows from a docking pose or an equilibrium constant. Kinetics is also the part of the problem that needs far more data than structural biology has produced: the field has been working with still pictures, and the next step is video.

Molecular dynamics (MD) simulation is the natural way to make that transition from pictures to video, and a growing family of trajectory resources has made simulation data reusable. MoDEL established shared trajectory libraries for soluble proteins[6]; GPCRmd provides a community repository for G-protein-coupled receptors[7]; MISATO released machine-learning-ready trajectories of protein–ligand complexes[8]; ATLAS collected 13.2 TB of trajectories describing protein flexibility[9]; DynaRepo extends simulation length to the microsecond scale[10]; PLAS-20k couples MD with solvation free-energy calculations[11]; and coarse-grained models probe allosteric landscapes at a fraction of the cost[12]. These resources differ in scope and force field, but most describe local relaxation around a bound pose and are validated by root-mean-square deviation (RMSD) from that pose. Because RMSD is bounded by the size of the binding site, a trajectory that never leaves the pocket can still appear converged. What such datasets sample is a quasi-static relaxation, not a dissociation event.

Complete dissociation trajectories change this. In DD-13M we followed 680 complexes from the PDBbind+ koff subset through 50 replicas each and retained 26,612 trajectories for 565 complexes—to our knowledge, the first large corpus of end-to-end unbinding events[13]. DD-13M showed that dissociation is simulable at scale; it also exposed the next limitation, which is analytical rather than computational. MDRepo[14] now holds more than 80 TB of such trajectories, yet the field has no settled way to use an archive of this size. The common responses are to compress a trajectory into a few static descriptors and fit koff, or to hand it to a generative model and treat the model as a black box that emits new trajectories. Neither turns the archive into something that can be compared, queried and reasoned over. What is missing is a method for tokenising a molecular dynamics database, and DD-03B is our attempt to build one.

We assembled 19,037 protein–ligand complexes from the freely available PDBbind+ v2020R1 set and generated complete dissociation trajectories with the DD-13M pipeline, obtaining 766,550 trajectories (290.6 million frames, 39.9 TB) for 15,540 complexes. Each complex is written as a uniform record of more than 100 descriptors, so that trajectories can be compared as tokens rather than as coordinates. Grouping the records by UniProt accession gives 3,687 accessions, a unit above the PDB entry, and lets us ask a question that structural biology cannot:

whether binding affinity is related to the pocket a ligand occupies and to the pathway by which it leaves. The resulting framework — UniProt-grouped, target-level pathway analysis — treats the ligands of a protein as a population rather than as independent anecdotes. We then take two representative, well-populated UniProt families — one with an open pocket and one with a closed, multi-pathway pocket — and use them to set out how each type of site should be analysed, before turning to the limits of the UniProt accession as a classifier.

## Results

### A corpus of complete dissociation events

Of the 19,037 complexes, 18,439 (96.9%) were modelled successfully and 15,540 yielded at least one productive trajectory that passed curation (Fig. 1a,c). Up to fifty independent metadynamics replicas per complex yielded 766,550 trajectories and 290,605,927 all-atom frames, corresponding to 39.9 TB of raw H5MD data. Trajectory endpoints were clustered on the protein surface and refined with nudged elastic band calculations, yielding 15,844 reproducible minimum free-energy pathways and 15,540 binding-pocket angiography (BPA) maps. Each pathway is stored as 100 milestone frames; each BPA is a four-dimensional free-energy volume that renders the spatial probability of ligand residence during escape. Relative to DD-13M, the number of complexes increased 28-fold, the number of trajectories 29-fold and the stored volume 45-fold; the full comparison is given in Fig. 1c, and the complete descriptor set is listed in the descriptor dictionary released with the dataset.

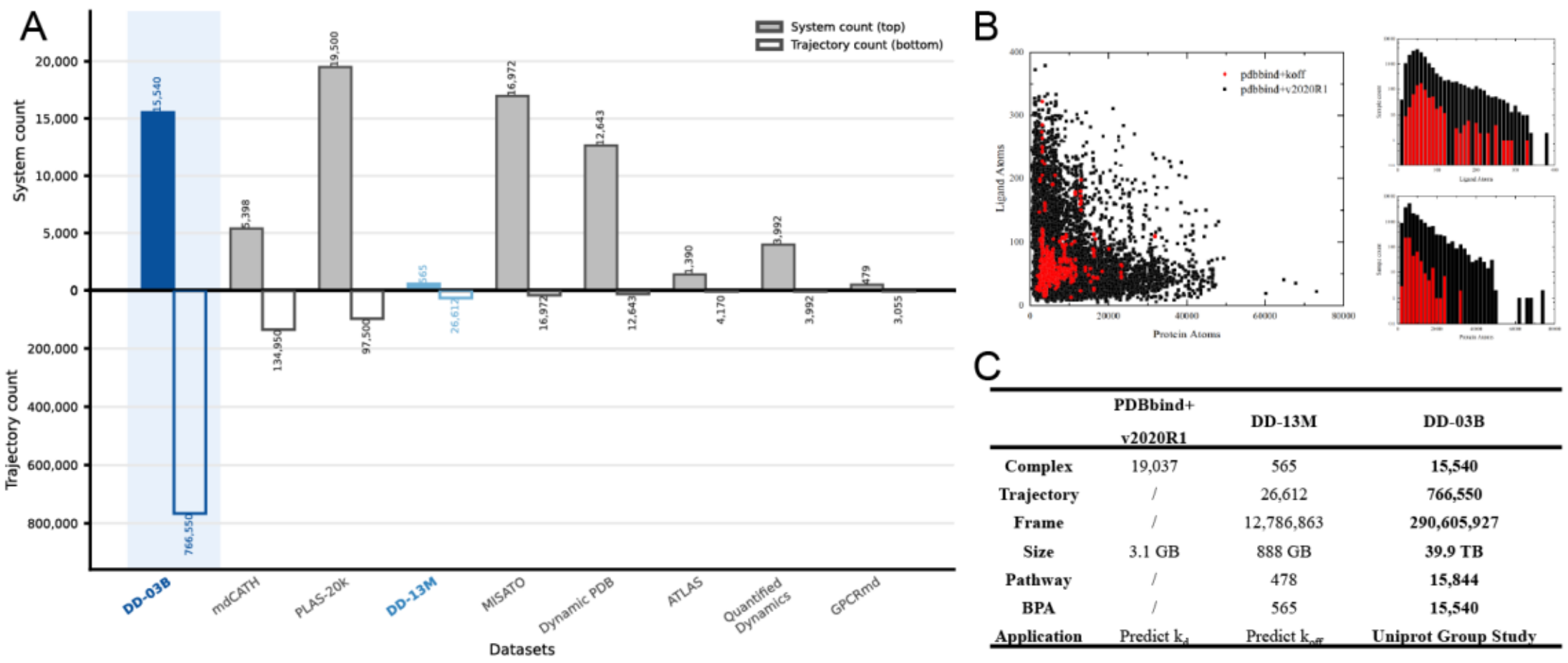

| | PDBbind+ v2020R1 | DD-13M | DD-03B |
|---|---|---|---|
| **Complex** | 19,037 | 565 | **15,540** |
| **Trajectory** | / | 26,612 | **766,550** |
| **Frame** | / | 12,786,863 | **290,605,927** |
| **Size** | 3.1 GB | 888 GB | **39.9 TB** |
| **Pathway** | / | 478 | **15,844** |
| **BPA** | / | 565 | **15,540** |
| **Application** | Predict $k_d$ | Predict $k_{off}$ | **Uniprot Group Study** |

**Fig. 1 | Composition of the DD-03B dataset. a**, System count (top) and trajectory count (bottom) for DD-03B and eight comparable trajectory resources; DD-03B contributes 15,540 systems and 766,550 trajectories, against 565 systems and 26,612 trajectories for DD-13M. **b**, Ligand heavy-atom count versus protein atom count for the 19,037 source complexes (black, PDBbind+ v2020R1 free subset used for DD-03B; red, PDBbind+ koff subset used for DD-13M), with marginal distributions of ligand and protein atom counts. **c**, Comparison of PDBbind+ v2020R1, DD-13M and DD-03B: complexes, trajectories, all-atom frames, stored volume, curated pathways, BPA maps and intended application.

The corpus also broadens the covered chemical space. The largest ligand contains 380 atoms (322 in DD-13M) and the largest protein 73,176 atoms (31,855 in DD-13M); the joint distribution of ligand and protein size is shown in Fig. 1b, with the DD-13M koff subset and the DD-03B source set distinguished. DD-03B extends the joint distribution to larger complexes in both dimensions than the DD-13M subset, so the corpus covers ligand–target pairs that were previously undersampled.

Not every complex yields a usable dissociation event, and the failure modes are informative. Within the 15,540 systems that carry trajectories, manual setup was required for 15.2%, mostly because of non-standard residues or unusual ligand chemistry. A further 3.4% fell in a shallow-pocket class: the ligand already sits at the protein surface and satisfies the escape criterion within the first frame, so these entries were not counted as dissociations and were excluded from all pathway statistics below. Short-pathway systems (18.1%) were simulated successfully but produced pathways shorter than 5 Å, indicating that the ligand was outside a well-defined pocket. Single-pathway systems accounted for 47.1% and multiple-pathway systems for 16.3%. Thus about one in four complexes with a resolved pathway uses more than one reproducible exit pathway—the first indication that a single dissociation pathway is the exception rather than the rule.

## A uniform record makes dissimilar pathways comparable

Comparing pathways across ligands requires a description that is richer than RMSD but still uniform across systems. We therefore store, for every modelled complex, a single record of 24 static fields and 95 dynamic fields, organised into five dimensions (Fig. 2). The static half describes the system: PDB and UniProt identifiers, ligand name, formula, InChI, molar mass, atom count and long-axis length; protein residue and atom counts and molar mass; the initial protein–ligand centre-of-mass distance; the size of the protein (radius of gyration, solvent-accessible surface, sphericity and three volume definitions); and the space accessible to the ligand. The dynamic half describes what the trajectory did: the system tag, atom count, valid frames, replica count and file sizes; the mean, dispersion and range of ligand RMSD, radius of gyration, protein–ligand centre-of-mass distance and clash score; the accumulated bias work; the escape volume (pocket volume, hit-zone volume and sphericity, ligand-cloud volume and sphericity); and, for each resolved pathway, its distance, exit angle, share and replica support, together with the summary path metrics. Each dimension is a fixed set of quantities with fixed units, so a record from one complex can be placed directly beside a record from another. In effect, the record tokenises a trajectory: a variable-length set of frames becomes a fixed-length description that can be compared, joined and searched.

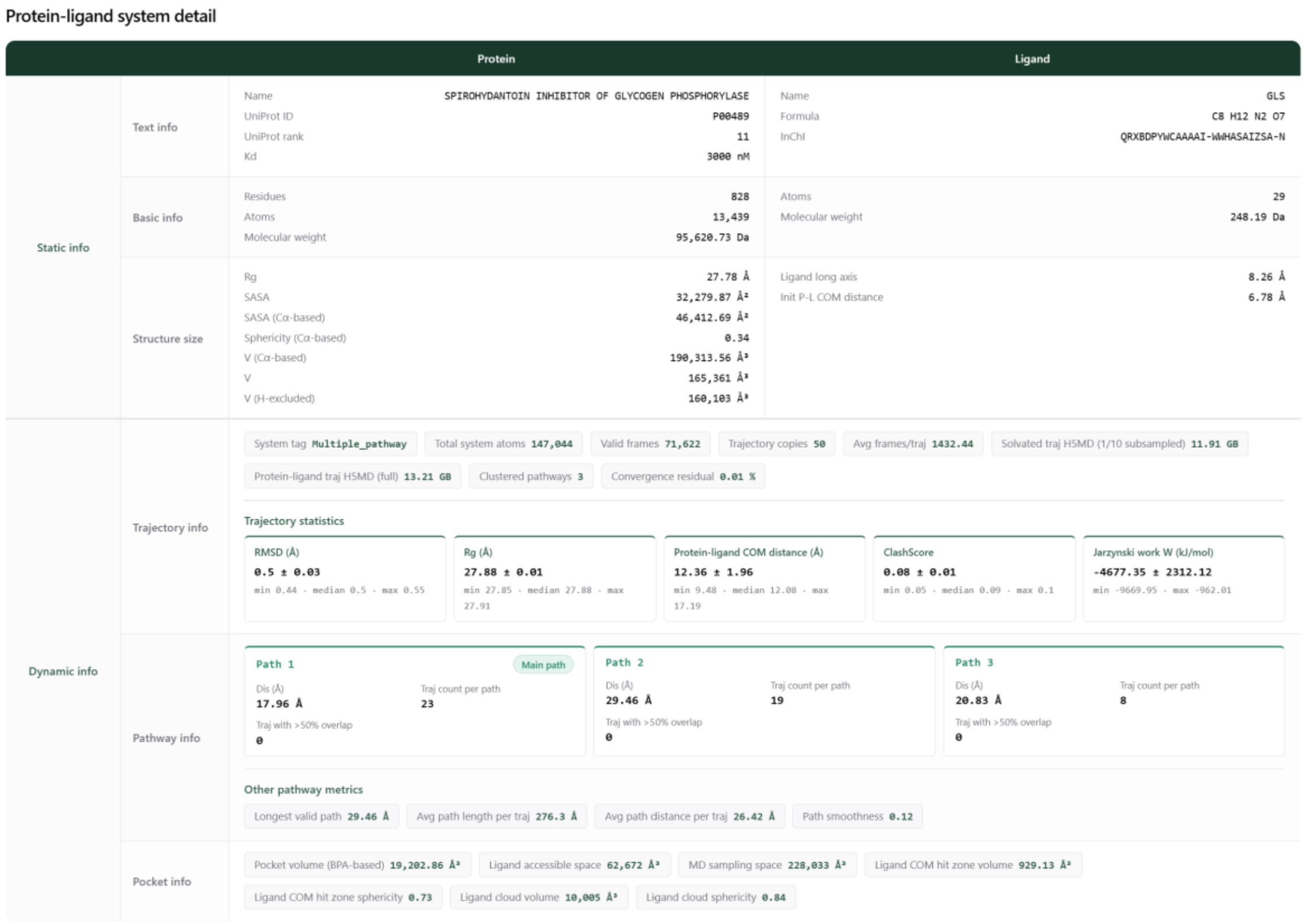

**Fig. 2 | System-level annotation record.** The record integrates five dimensions for a single complex (shown for P00489 with ligand GLS). **Text and basic information** gives protein and ligand identity, residue and atom counts and molecular weight. **Structure size** gives radius of gyration, solvent-accessible surface area (total and Cα-based), sphericity, molecular volume (all-atom, Cα-based and hydrogen-excluded), ligand long-axis length and the initial protein–ligand centre-of-mass distance. **Trajectory statistics** gives ligand RMSD, radius of gyration, protein–ligand centre-of-mass distance, clash score and the accumulated bias work. **Pathway information** gives the number of clustered pathways, the path distance, replica count and >50% overlap count of each pathway, the longest valid path, the mean path length and mean path distance per trajectory, path smoothness and the convergence residual. **Pocket information** gives the BPA-based pocket volume, the ligand-accessible space, and the volume and sphericity of the centre-of-mass hit zone and of the ligand cloud.

The record is meant to be read as a page rather than queried field by field. The static and dynamic halves of a complex, its trajectory statistics, its per-pathway breakdown and its escape volume sit side by side (Fig. 2), and the pathway block reports how many pathways were resolved, how many replicas support each, and how far and at what angle each pathway leaves the site. Records are versioned with a provenance entry (software version, force field, seed range and completeness flag), and missing values are retained as explicit nulls rather than imputed, so a user can select the subset of descriptors and systems for which a comparison is valid instead of inheriting someone else's imputation.

## The target-level landscape of exit multiplicity

Grouping the 19,037 source complexes by UniProt accession produced 3,687 groups (Fig. 3a). The size distribution is heavy-tailed, as expected for a resource built from the Protein Data Bank[15]: the largest group contains 425 structures and the count falls below 200 by the fourth rank, while most groups contain fewer than five. This skew is the statistical substrate of the analysis. Within a large group the protein is approximately fixed and the ligand varies, so variation in dissociation behaviour can be attributed to the ligand or to protein–ligand coupling rather than to a different target.

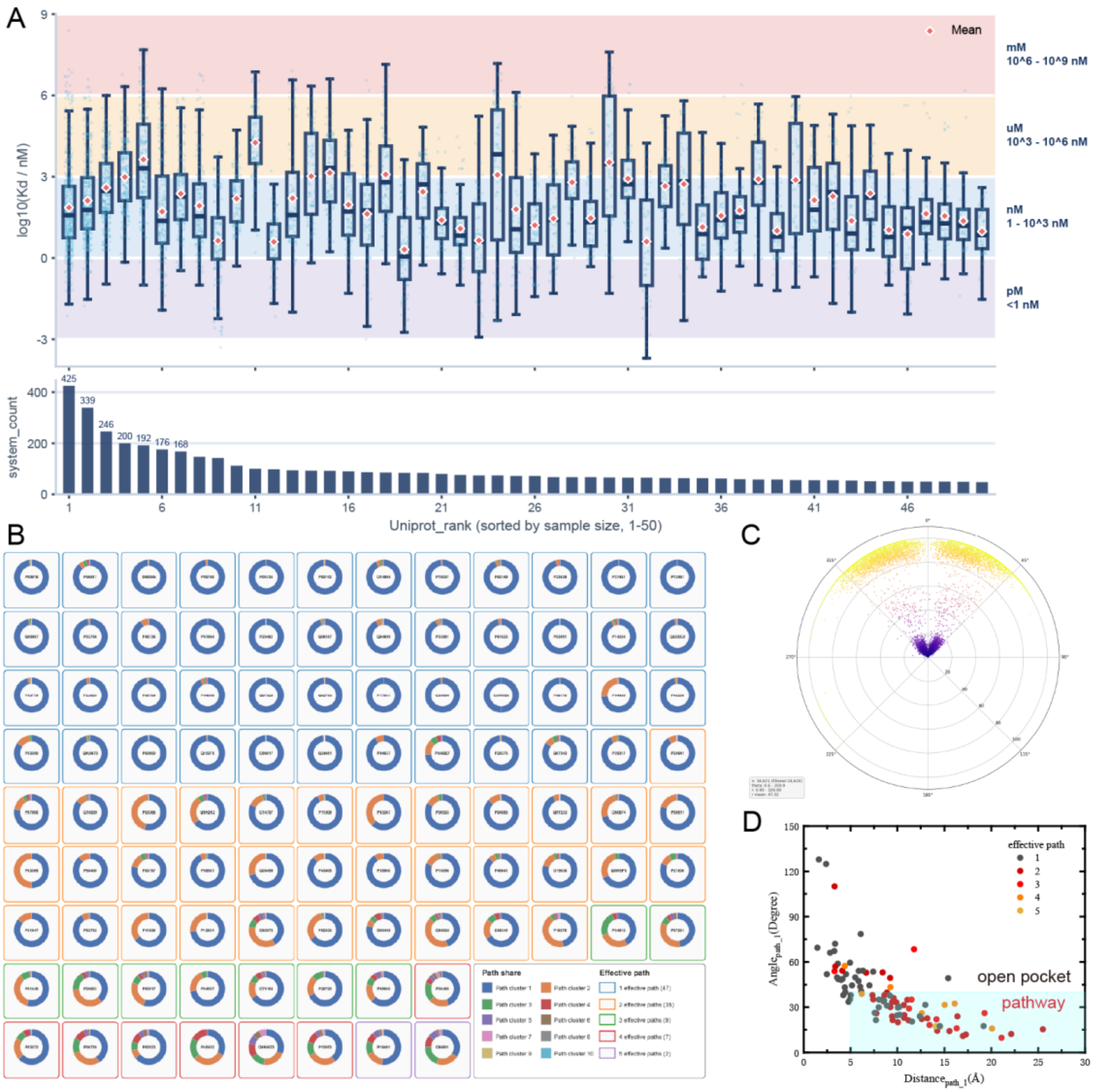
A
Mean
mM
10^6 - 10^9 nM
uM
10^3 - 10^6 nM
nM
1 - 10^3 nM
pM
<1 nM
log10(Kd / nM)
system_count
425
339
246
200
192
176
168
Uniprot_rank (sorted by sample size, 1-50)
B
Path share
Effective path
C
D
effective path
open pocket
pathway
Angle$_{path_1}$(Degree)
Distance$_{path_1}$(Å)

**Fig. 3 | The UniProt-grouped landscape. a**, Group-size distribution (lower panel) and experimental affinity (upper panel) for the 50 largest groups, ordered by sample size. Boxes show the median and interquartile range of log10(Kd/nM), whiskers extend to 1.5× the interquartile range, diamonds mark the mean, and the background bands delineate the pM (<1 nM), nM (1–$10^3$ nM), µM ($10^3$–$10^6$ nM) and mM ($10^6$–$10^9$ nM) ranges. **b**, Pathway composition of the 100 most populated UniProt groups; each sector gives the share of replicas assigned to a pathway, and the box colour gives the number of effective pathways (1, n = 47; 2, n = 35; 3, n = 9; 4, n = 7; 5, n = 2). **c**, Polar summary of trajectories relative to their cluster mean. The angle is the angle between a trajectory and the mean trajectory of its cluster; the radius measures the deviation of that trajectory from the cluster mean, so that a larger radius means lower overlap. The informative part is the angular spread near the centre, where overlap is high. **d**, Angle of the dominant pathway (AngPath_1) versus its distance from the binding site (Distance_path_1), coloured by the number of effective pathways. The shaded region marks the pathway-dominant call used for triage: Distance_path_1 > 5 Å and Angle_path_1 < 40°. The cut is a guide rather than a classifier.

The spread of experimental Kd within a single group is frequently several orders of magnitude, comparable to the spread across the entire database (Fig. 3a). An affinity value is therefore not a property of a target but of a target–ligand pair, and the group is the correct unit within which to interpret it.

The same logic applies to pathway descriptors, and here the record already carries them in a form that can be compared across groups. For each group we plot the share of sampled replicas assigned to each clustered pathway as a circular composition diagram (Fig. 3b). Across the 100 most populated groups, 47 use a single effective pathway, 35 use two, 9 use three, 7 use four and 2 use five. The composition diagrams make the difference visible at a glance: groups that are dominated by one sector sit next to groups in which three or four sectors of comparable size coexist. Pathway multiplicity is therefore not a property of the database as a whole but a property of individual targets, and within a target it can differ from ligand to ligand.

The spatial counterpart of this composition is the geometry of the first exit. Figure 3c summarises how tightly a trajectory follows its own average pathway: the angle is the angle between a trajectory and the mean trajectory of its cluster, and the radius is the deviation of that trajectory from the cluster mean, so that the high-overlap trajectories sit near the centre and their angular spread is the part that carries information. Plotting the angle of the dominant pathway against its distance from the binding site then separates two families of systems (Fig. 3d). Complexes with one or two effective pathways occupy the short-distance, high-angle region: the ligand leaves close to the pocket mouth and in directions that spread over a wide arc. Complexes with three to five effective pathways extend to longer path distances and smaller exit angles, that is, to pathways that follow a defined channel away from the site. The shaded region of Fig. 3d marks this division: a complex is called pathway-dominant when its dominant pathway is longer than 5 Å and leaves at an angle below 40°. The cut is a guide to reading the plane rather than a classifier, and the complexes near its edges are the informative ones. The same two coordinates therefore carry both the geometric classification (open versus channelled) and the multiplicity classification (single versus multiple pathways), and they can be computed for every curated complex.

### Docking-pose determinism holds for an open pocket

P00918 is the most abundant group in the database (425 structures) and a clear illustration of the case in which the classical picture works. Of 390 systems with trajectories, 389 were structurally alignable and 381 were retained in the main pocket after exclusion of surface-bound ligands, drawn from 19,027 trajectories (Fig. 4a). The site is an open, funnel-shaped cavity: wide and shallow at the mouth, with no single channel to follow, and the initial ligand binding positions spread across the pocket mouth rather than concentrating in a single subsite (Fig. 4b). The pathway composition is correspondingly dominated by one sector: 97.9% of replicas fall into the first pathway (Fig. 4c), while the trajectories scatter around their cluster mean without breaking into separate directional groups (Fig. 4d). In such a site, dissociation is less a pathway than a departure: the ligand leaves as soon as it moves a little away from its docking pose, the exit directions do not separate into channels, and there is no single path to define. A shell-like radial distance, rather than a path, is the natural coordinate.

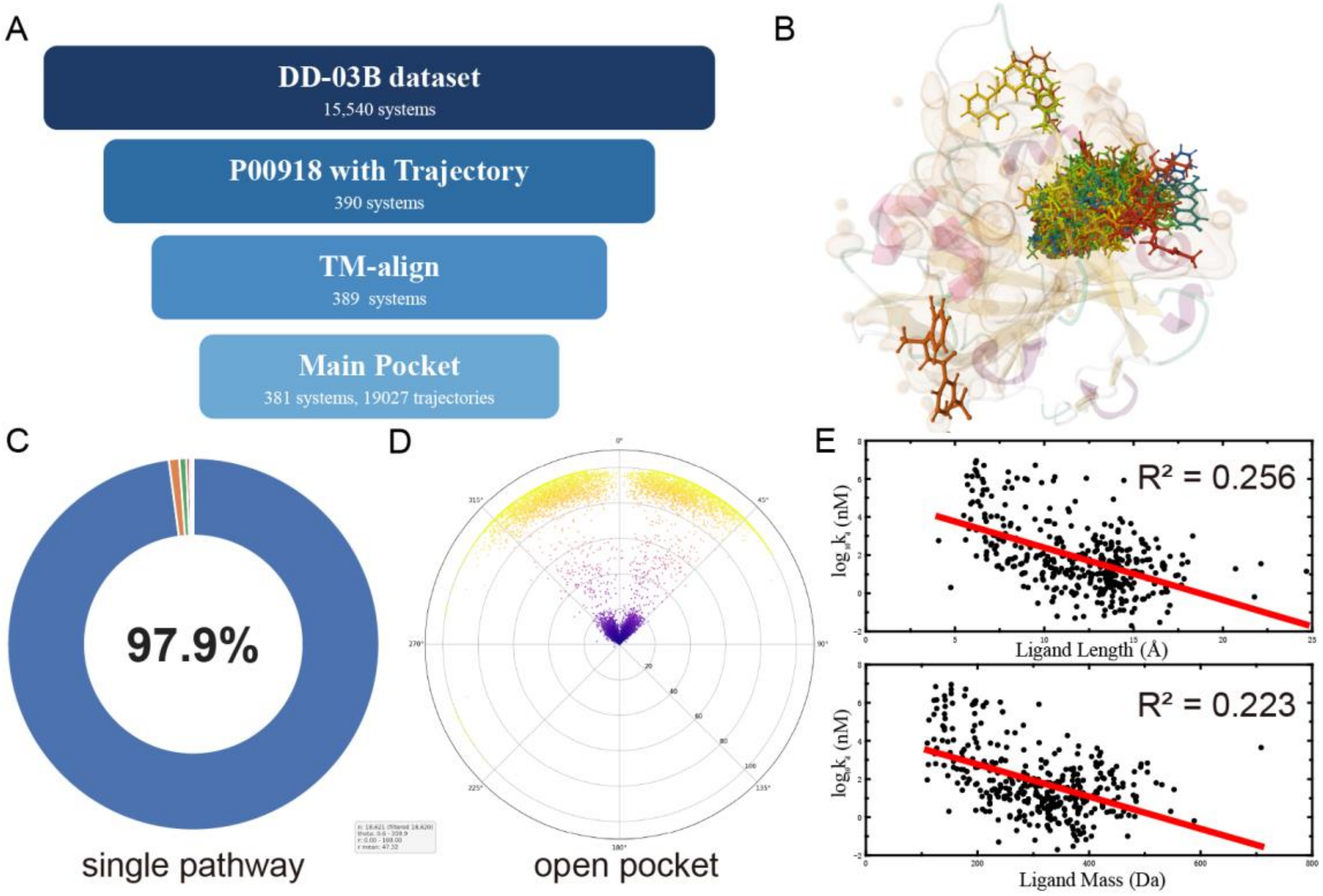

**Fig. 4 | P00918: an open, funnel-shaped pocket. a**, Group workflow: 15,540 systems in DD-03B → 390 P00918 systems with trajectories → 389 structurally alignable → 381 retained in the main pocket (19,027 trajectories). **b**, Initial ligand binding positions, spread over the pocket mouth, coloured by ligand-size class. **c**, Pathway composition: 97.9% of replicas fall into a single pathway. **d**, Polar summary of trajectories relative to the cluster mean, as in Fig. 3c: the angle is the angle between a trajectory and the mean trajectory of its cluster, and the radius is the deviation from that mean, so that a larger radius means lower overlap. The high-overlap trajectories form one continuous cloud around the mean direction rather than separate directional groups, which is what an open pocket means operationally in this dataset. **e**, log10 Kd versus ligand long-axis length (top; $R^2$ = 0.256) and versus ligand mass (bottom; $R^2$ = 0.223); points are individual complexes and the red line is the ordinary least-squares fit.

The group is also a reminder that the labels are ours, not the data's. A reader wants a tag — open or pathway, single or multiple — but the underlying distributions are continuous and any boundary is drawn by hand. Figure 3d places that boundary by eye, and P00918 falls just inside it—a little beyond 5 Å in distance and a little below 40° in angle—so it is labelled pathway, although its pathway composition and exit fan say open. It is a borderline case between the two classes rather than a canonical member of either, which makes it a useful calibration for where the boundary sits. Rather than insist on a binary tag, we report the quantitative tokenised record and let the reader place the cut-off.

Affinity in this group behaves the way a docking score assumes. Across the 381 complexes, log10 Kd decreases with ligand long-axis length ($R^2$ = 0.256) and with ligand mass ($R^2$ = 0.223; Fig. 4e): larger ligands make more contacts and bind more strongly. The trend is what is expected when affinity is carried by the contacts a ligand makes: contact area grows with size, and within a chemically similar series the functional groups on offer change little. What matters here is that a single well-defined pose carries the information, because there is no second mode for an ensemble to average over. A thermodynamic decomposition of this group is a natural next step, but a size–affinity correlation alone does not establish how much of the affinity is enthalpic and how much entropic, and we do not attempt that attribution here. The correlation is modest because ligand size is only a proxy for contact quality, but its direction is unambiguous. For an open pocket, therefore, a pose plus a size is a sufficient description of binding: the pose-deterministic picture is not wrong, it is the correct limit for this class of site.

## When does the exit ensemble matter?

If the open-pocket picture generalised, the exit-ensemble term would stay negligible and a pose plus a size would always suffice. The opposite limit is easy to specify: a confined pocket with several reproducible exits. P00489, the eleventh largest group, is the clearest case, and the reason a pose is not always enough. Ninety-eight of its 100 structures yielded trajectories, 89 were alignable, and 56 systems produced 2,800 trajectories in a confined, close pocket (Fig. 5a,b). The same 56 systems resolve into seven exit pathways with markedly different weights — C1, 44.3%; C2, 18.6%; C3, 17.4%; C4, 5.6%; and 14.0% over the remaining pathways — and the per-ligand composition diagrams are not copies of one another: some ligands put almost all replicas in one channel, others spread probability over three or four (Fig. 5c). The three-dimensional reconstruction shows that these are spatially distinct channels leaving the same site (Fig. 5d).

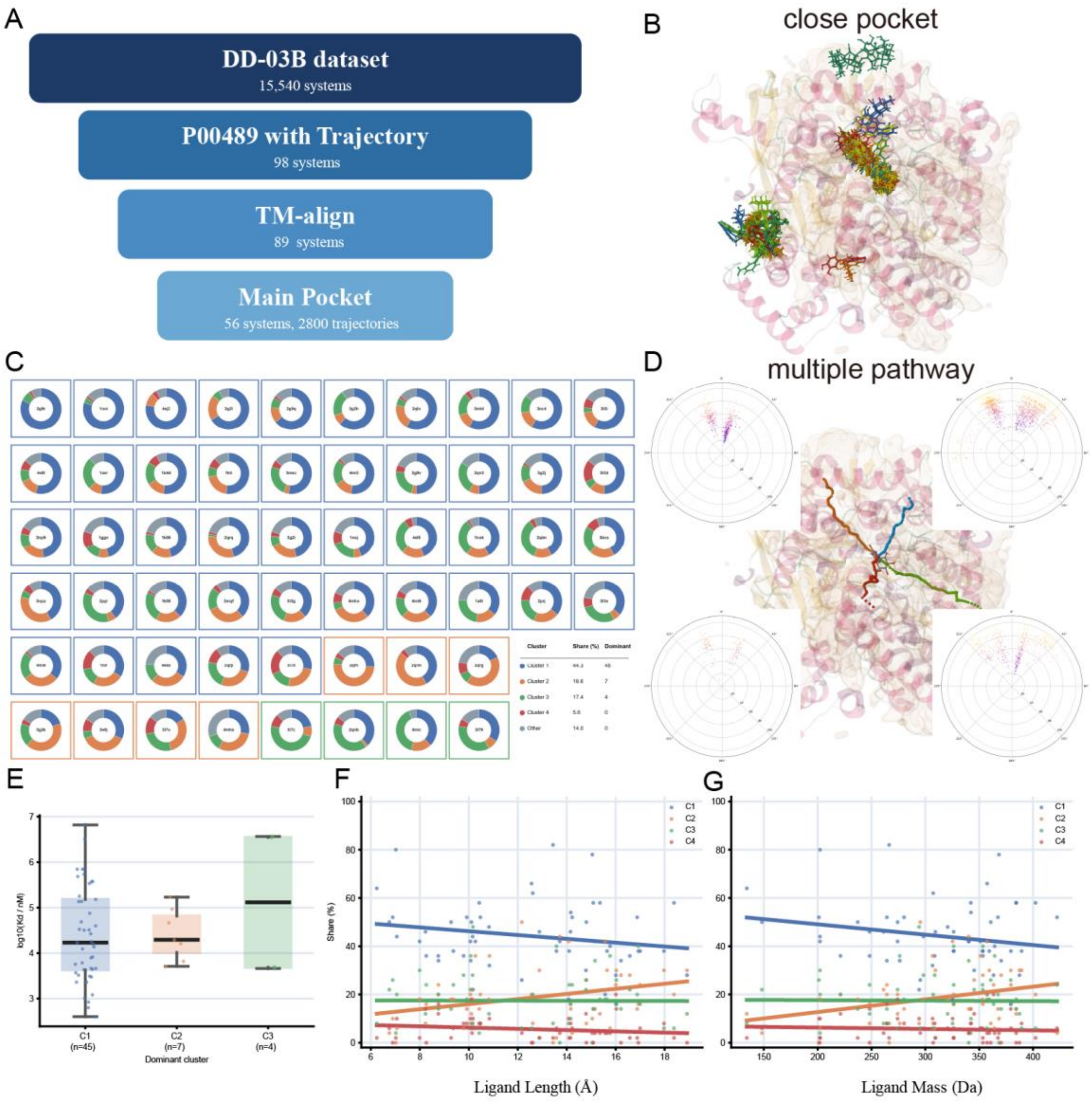
A
DD-03B dataset
15,540 systems
P00489 with Trajectory
98 systems
TM-align
89 systems
Main Pocket
56 systems, 2800 trajectories
B
close pocket
C
Cluster
Share (%)
Dominant
Cluster 1 44.3 45
Cluster 2 18.6 7
Cluster 3 17.4 4
Cluster 4 5.6 0
Other 14.0 0
D
multiple pathway
E
log10(Kd / nM)
C1 (n=45)
C2 (n=7)
C3 (n=4)
Dominant cluster
F
Share (%)
Ligand Length (Å)
C1
C2
C3
C4
G
Ligand Mass (Da)
C1
C2
C3
C4

**Fig. 5 | P00489: ligand-specific exit pathways. a**, Group workflow: 98 systems with trajectories → 89 alignable → 56 systems in the main pocket (2,800 trajectories). **b**, Ligand poses in the confined, close pocket. **c**, Pathway composition for the 56 ligands; the inset gives the group-level share of each cluster (C1, 44.3%; C2, 18.6%; C3, 17.4%; C4, 5.6%; other, 14.0%) and the number of ligands for which each cluster is dominant (C1, 45; C2, 7; C3, 4). **d**, Three-dimensional reconstruction of the multiple exit pathways (coloured paths) over the protein surface. The four polar insets give, for representative ligands, the same trajectory-versus-cluster-mean summary as Fig. 3c; from top-left to bottom-right they correspond to C2 (orange), C1 (blue), C4 (red) and C3 (green), using the cluster colours of Fig. 5c. Each ligand shows separate angular groups, consistent with distinct channels. The same colour code is used for the cluster shares in f and g and for the Kd boxes in e. **e**, log10 Kd grouped by dominant pathway (C1, n = 45; C2, n = 7; C3, n = 4); boxes show the median and interquartile range and whiskers extend to 1.5× the interquartile range. **f**, Share of each cluster versus ligand length. **g**, Share of each cluster versus ligand mass. Lines in f and g are ordinary least-squares fits.

Here the second term cannot be ignored. The number of exits available to a ligand, and the probability it assigns to each, enters the binding partition function as a sum over exit channels rather than as the depth of one, and it is the hardest term to compute: it converges slowly, it depends on how many channels are resolved, and a free-energy estimate along one channel says nothing about the alternatives. Fifty independent escapes per complex give an empirical handle instead. Rather than assume the distribution, we estimate it: the vector of per-pathway shares is read as an estimate of the exit-ensemble probabilities, and the ensemble itself—rather than the dominant path—becomes the object of analysis. We do not convert this vector into a numerical entropy. With 50 replicas the estimator is not stable enough to be quoted as a thermodynamic quantity, and we therefore treat it as a proxy for the pathway term of the partition function rather than as a measured entropy. To our knowledge this is the first systematic dissection of dissociation at the level of the exit ensemble, and it is what makes the ligand dependence visible at all.

The ensemble reorganises with ligand size in a consistent direction. Small ligands concentrate their replicas in C1; as ligand length or mass increases, the C1 share falls and the C2 share rises, while the C3 and C4 shares decline more slowly (Fig. 5f,g). The protein is held fixed throughout, so the change is a property of the ligand. A natural reading is steric: a larger ligand pays more to pass through the C1 channel and redistributes into C2. What kind of statement is this? Hindrance raises the cost of one channel relative to another, which is a local interaction term; what changes is how the ligand distributes over the channels that are available, that is, the degeneracy of the exit ensemble. The switch is therefore carried by the shape of the exit landscape rather than by binding strength, and it need not be accompanied by any change in the depth of the bound basin. Whatever the microscopic cause, the observable is robust: the same pocket is left by different pathways depending on how large the ligand is.

If pathway selection were simply another readout of binding strength—stronger ligands leaving through higher barriers—then the dominant pathway and the affinity would be tightly correlated. They are not. Within P00489, ligands grouped by dominant pathway have broadly overlapping Kd distributions (Fig. 5e; C1, n = 45; C2, n = 7; C3, n = 4). The median of the C1 subset is the lowest of the three, but the ranges overlap almost completely, and the C2 and C3 subsets are too small for a quantitative comparison. The median Kd moves by about one log10 unit from C1 to C3 (Fig. 5e). That is not nothing—a tenfold difference in affinity is larger than the scatter between the assay types pooled in PDBbind—but it is small against the several orders of magnitude spanned within each subset, and against the near-complete change in pathway composition over the same complexes. Pathway identity therefore carries little information about affinity, and the two quantities are at most weakly coupled. A tighter statement would need more ligands per class or a variance decomposition across groups, and this group supplies neither.

This weak coupling has a mechanistic interpretation. Equilibrium affinity is an integral over the bound basin; if several pathways have comparable barriers, the depth of the basin is insensitive to which pathway dominates. Kinetics are not symmetric to this averaging. The barrier along the dominant pathway sets the escape rate, so two ligands with the same Kd can, in principle, have different dissociation rates if their pathway distributions differ. We therefore read the weak Kd–pathway relation as evidence that affinity and pathway selection report on different features of the free-energy landscape, not as evidence that pathways are irrelevant. Testing this prediction requires measured rates: experimental koff data are being acquired for a subset of the systems analysed here, and we plan to incorporate koff fitting into a future release of DD-03B.

## Beyond the UniProt accession

Grouping by UniProt accession is the natural first step, and it is the unit we use throughout this study. An accession is a subspecies-level classification: it sits above the PDB entry, groups structures that share a sequence, and is the natural first unit for a target-level analysis. It is not the natural final unit. The object a pathway analysis needs is a structural class: a set of structures that share a pocket and a repertoire of exits. The two coincide often enough for accession-level grouping to be useful, and they diverge in two reproducible ways, both shown in Fig. 6. Here we characterise the exceptions and the handling they require.

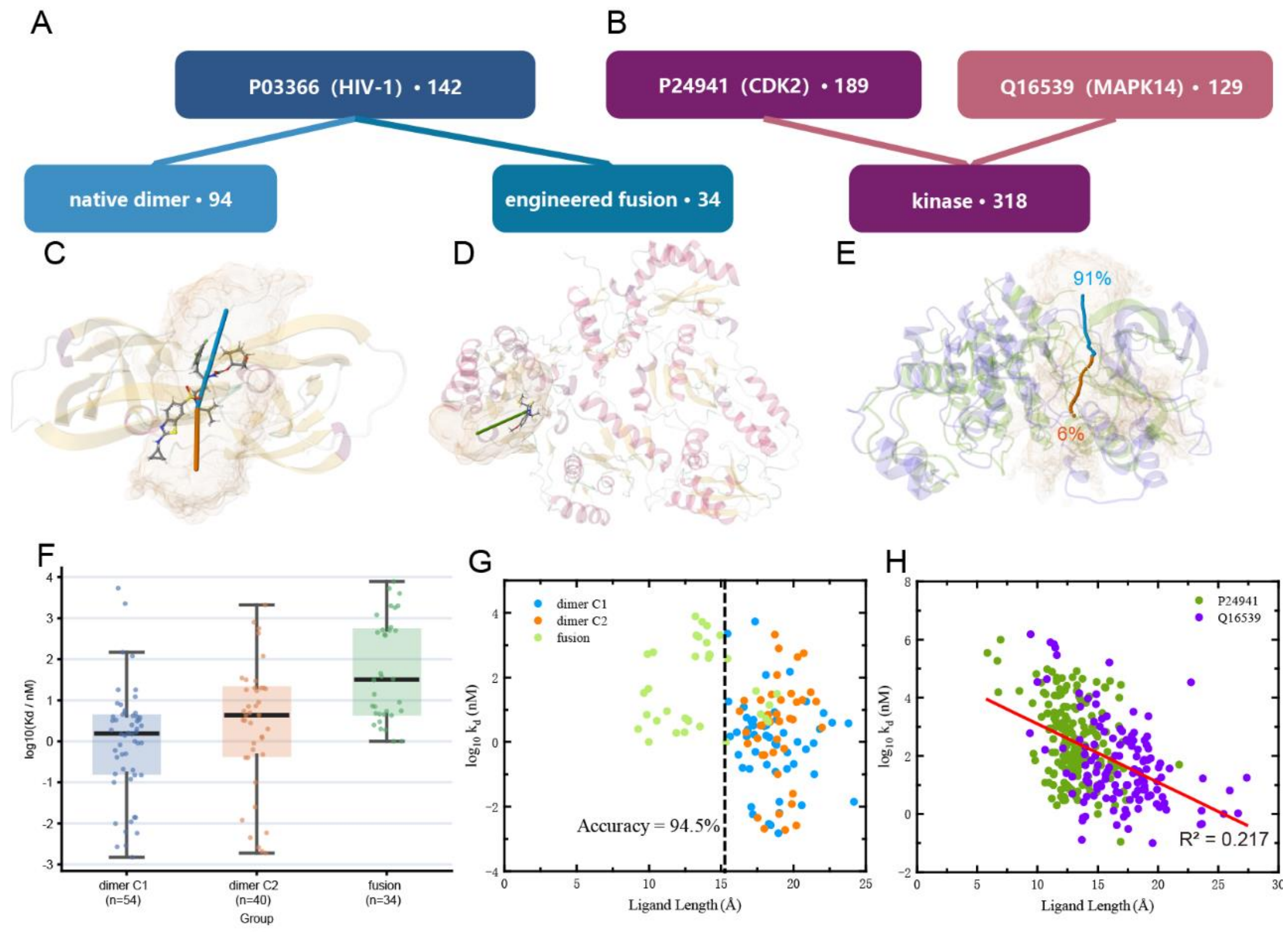

I

| Pattern | Groups | Uniprot | Example | Handling |
|---|---|---|---|---|
| **Domain Split** | 3 | 3 | HIV-1 (94:34) | split, analyze major class |
| **Degeneracy** | 12 | 51 | kinase (19 uniprot) | group-level comparison |
| **Auxiliary** | 1 | 1 | P0ABE7 / BRIL | exclude |

**Fig. 6 | Construct boundaries and accession degeneracy. a**, One-to-many (1→N): P03366 (HIV-1, 142 structures) splits into a native dimeric class (94 structures) and an engineered fusion class (34 structures) with different domain boundaries. **b**, Many-to-one (N-in-1): P24941 (CDK2, 189 analysed structures) and Q16539 (MAPK14, 129 analysed structures) converge on a single kinase class of 318 structures. **c**, **d**, Escape pathways of the dimeric and fusion classes. **e**, Superposition of the two accessions, with one pathway carrying 91% of replicas and a second 6%. **f**, log10 Kd by group (dimer C1, n = 54; dimer C2, n = 40; fusion, n = 34). **g**, log10 Kd versus ligand length for the P03366 classes; the dashed line marks the 15 Å threshold that separates the fusion class from the dimer classes with 94.5% accuracy. **h**, log10 Kd versus ligand length for P24941 and Q16539, with a common fit ($R^2$ = 0.217). **i, Summary of the three patterns.**

The first exception is one-to-many (1→N): one accession, several structural classes. Because an unalignable structure is itself a separate class, the number of splits grows without bound unless the minor class is required to be substantial; we therefore count a split only when the minor class contains at least 30 structures. Three accessions pass that threshold (P24941, 199:30; P03366, 94:34; P04585, 38:32); Fig. 6a shows P03366 (HIV-1). Of its 142 structures, 94 form a native dimeric class and 34 an engineered fusion class with different domain boundaries. The two classes differ in the properties that matter for design: the dimeric site binds more strongly (median log10 Kd ≈ 0.3–0.4 versus 1.5) and its ligands are larger (15–23 Å versus 10–15 Å), so that a single ligand-length threshold of 15 Å separates them with 94.5% accuracy (Fig. 6f,g); the escape geometry differs as well (Fig. 6c,d). One accession therefore carries two mechanisms, two affinity classes and two ligand-size classes. The handling is to split before pathway statistics and analyse the major class, while recording the split in the family table.

The second exception is many-to-one (N-in-1): distinct accessions, one structural class. Twelve degenerate groups cover 51 of the top-100 accessions, and the largest is the protein-kinase group, in which 19 accessions (1,293 structures) share an ATP-site architecture. We took the two most abundant members, P24941 (CDK2, 189 analysed structures) and Q16539 (MAPK14, 129 analysed structures), which form a single kinase class of 318 structures (Fig. 6b). They share their dominant pathway — 91% of replicas in the joint superposition and 6% in a second channel, with separate values of 90.8%/7.8% and 86.0%/9.2% for the two accessions (Fig. 6e) — and their affinity–size relations coincide, with log10 Kd falling along one line ($R^2$ = 0.217; Fig. 6h). For the purpose of pathway analysis these behave as one protein, and Fig. 6 works this through on a single pair as the template for a degenerate class.

The two annotations are independent: P24941 is also the representative of the kinase degeneracy group, so an accession can be a domain-split case and an N-in-1 anchor at once, and the family table records the roles in separate fields. One further entry (P0ABE7, BRIL) is an auxiliary binder rather than the annotated protein and is excluded. The three patterns are summarised in Fig. 6i. Splitting every 1→N case and merging every N-in-1 case at database scale requires structural clustering of all 19,037 structures and will be reported separately. Until then, the family table records the exceptions explicitly, so that a user can choose the unit appropriate to the question.

Defining a better accession, however, is not simply a matter of clustering more structures. Structural classification has a long history of this problem: CATH, SCOP and ECOD assign stable identifiers to domains and superfamilies and are the closest existing analogue to a structural accession[44–46], but they classify domains and folds, not the pocket-level classes a pathway analysis needs, and they illustrate two difficulties that any structural accession inherits. The first is the threshold: whether two structures belong to the same class depends on where the cut-off is placed, and a cut-off that is appropriate at 90% identity need not be appropriate at 95%; both the number of classes and the side on which a given structure falls change with the threshold. The second is that similarity is not transitive. A can be similar to B and B to C while A and C fall below the cut-off, so “same class” is not an equivalence relation and the partition depends on the order in which structures are merged. An accession that replaced UniProt would therefore have to specify the metric, the cut-off and a rule for resolving these chains, and none of the three is currently standardised. That is why we raise the question rather than answer it: the present results show that the accession is not the right unit and that the right unit has to be structural, but they do not settle how it should be drawn.

## Discussion

We have described a target-level framework for ligand dissociation and applied it to 766,550 complete trajectories spanning 15,540 complexes and 3,687 UniProt groups. The contribution is not a larger archive but a change of object. Instead of asking how one ligand leaves one protein, the framework asks how the ligands of a protein distribute their exits, and it treats that distribution as a property of the target–ligand population. That shift makes two things measurable which a single trajectory cannot show: whether exit-pathway selection is a property of the ligand, and whether affinity predicts it.

Pocket and ligand act on the same quantity at different levels. In open, shallow sites the repertoire of exits collapses to one broad mode, affinity follows ligand size and the extent of contact, and a docking pose is a sufficient description, because there is only one exit to choose. In confined sites the same protein resolves several exits, ligands of different size weight them differently, and affinity no longer predicts the pathway: the dominant pathway and the Kd distribution overlap almost completely, and the share of a pathway changes with ligand size while Kd changes only weakly. The exit-ensemble term is negligible in the first case and dominant in the second, which is why, in the second, the ensemble and not the dominant path has to be the object of analysis.

The framework also shows that the unit of analysis is not given by the database. UniProt accession is a workable first grouping, but it is sequence-based, while the object a pathway analysis needs is a structural class. The two diverge in both directions: one accession can hold two mechanisms with different affinities and different ligand-size classes, and distinct accessions can hold one, sharing a dominant pathway and a single affinity–size relation. We handle the first by splitting before pathway statistics and the second by group-level comparison. We do not claim a replacement accession, because defining one would require clustering the whole database and would still have to fix a similarity threshold and a rule for non-transitive chains; that is the question we leave open.

These results connect to, and extend, earlier work on multiple unbinding pathways. Multidirectional exit has been described for individual systems and exploited in path-based and steered simulations[16–21], and GPCR dynamics has shown how functionally important pathways can differ between ligands[22]. What has been missing is a population of sufficient size to ask whether pathway selection is a general ligand property and whether it carries information beyond affinity. The present dataset provides that population. It also provides the negative result that makes the positive one interesting: if pathway selection merely tracked affinity, it would add nothing to a Kd. It does not track affinity, so pathway selection is a candidate independent axis of ligand behaviour—candidate, because independence from one variable is not yet evidence of information. That step needs measured koff, and it is the test we are setting up.

The framework has practical implications. For simulation design, it argues for a triage step before expensive enhanced sampling: exit angle, path distance and the number of effective pathways are cheap descriptors that indicate whether a channel-based or a pocket-centric treatment[23,24] is the better starting point. For medicinal chemistry, it argues that construct choice is part of lead selection, not an artefact of crystallography, and that two compounds with the same pose may not have the same residence time. For database design, it argues that a trajectory archive becomes scientifically useful only when it is annotated densely enough to support comparisons across ligands; coordinates alone cannot answer target-level questions.

Several limitations should be stated. The production runs are short (2 ns per replica) and use metadynamics to accelerate escape; the trajectories are physically plausible and internally consistent, but they are not unbiased kinetic trajectories. The metadynamics bias distorts absolute timescales, so escape times and any rate-like quantity derived from individual trajectories are not interpreted kinetically; pathway shares are relative measures under a common protocol, and a share near 50% carries a binomial standard error of about 7 percentage points at 50 replicas, so the exit-ensemble probabilities estimated here are proxies rather than converged thermodynamic quantities. The stored bias work is exploratory: no free-energy estimate is formed from it, and it is not used to support any conclusion in this study. The affinity annotations inherited from PDBbind+ mix Kd, Ki and IC50, so group-level affinity comparisons are trends rather than precise thermodynamic quantities. Subgroup sizes within P00489 are small (n = 4–7 for the C2 and C3 subsets), so the comparison across dominant pathways is descriptive; we report distributions rather than subgroup tests. The three patterns in Fig. 6 are reported qualitatively for the minority classes; systematic structural clustering of the database is future work.

Our study is descriptive in its kinetic claims, and three questions follow from it. Can the pathway ensemble be predicted from a structure without simulating it? Does the pathway share carry information about residence time that Kd and pose do not? And can the grouping be made automatic, so that structural classes rather than accession strings define the unit? If the answers are yes, target-level pathway analysis becomes not only a way to describe dissociation but a way to design for it. The implication of the present results is narrower and firmer: the mechanism is not a property of the pose alone. For open sites the pose is a good proxy; for confined sites the object is a distribution over exits, and that distribution is set jointly by the pocket and the ligand.

## Methods

### Dataset assembly and system preparation

Structures were taken from the freely available subset of PDBbind+ v2020R1, which contains 19,037 experimentally resolved protein–ligand complexes annotated with binding data (Kd or IC50)[3–5] and deposited in the Protein Data Bank[15]. No filtering by affinity, protein family or ligand class was applied. Each complex was used as the starting conformation for an automated pipeline derived from DD-13M[13]. Proteins were supplied in PDB format and ligands in MOL2 format; systems that failed automatic construction were flagged for manual curation.

### Molecular dynamics protocol

Simulations were performed with SPONGE[25] using an automated pipeline for system construction, equilibration, production sampling and termination. Initial complexes were prepared with Xponge[26]. Proteins were described by AMBER ff14SB[27] and ligands by GAFF[28]. Each complex was solvated in a periodic TIP3P water box[29] with a minimum solute–box distance of 2.0 nm and neutralised with K+ and Cl−.

Energy minimisation used steepest descent, followed by 500 ps of NVT equilibration at 300 K (Langevin thermostat[30], coupling constant 1.0 ps) and 500 ps of NPT equilibration at 1 bar (Andersen barostat[31]). Heavy atoms of protein and ligand were restrained during equilibration. A further 1 ps NVT run assigned random velocities.

Production runs used metadynamics[32,33] with a 1 fs time step and the Cartesian coordinates of the ligand centre of mass as the collective variable[34]. Gaussian potentials with a fixed height $w = 2.5$ kJ/mol and width $\sigma = 0.1$ nm were deposited every 1 ps. In contrast to well-tempered metadynamics[35], the fixed-height variant was chosen because it accelerates escape and, when averaged over many independent replicas, yields a direct estimate of the three-dimensional free-energy surface. Protein Cα atoms were restrained throughout production. A solvent-accessible-surface criterion served as the reaction boundary: a simulation terminated as soon as the ligand centre of mass reached the protein SASA. Each production run was capped at 2.0 ns. Fifty independent replicas with different random seeds were run per complex, and coordinates were saved every 1.0 ps. Because the bias accelerates escape, absolute timescales are not physical. All comparisons of pathway selection are therefore made between systems simulated with identical bias parameters and replica counts, and are interpreted as relative probabilities rather than rates.

### Trajectory processing and pathway curation

For every complex, the endpoints of all successful replicas were projected onto the reaction surface and clustered in the plane tangential to the protein surface to identify exit channels[36]. Cluster representatives were refined with the nudged elastic band method[37], yielding 19,450 candidate minimum free-energy paths. Three filters were applied: paths shorter than 5.0 Å were discarded; paths with mean squared error above 200 $Å^2$ were discarded as non-convergent; and clusters supported by a single replica were discarded as non-reproducible. The curated set contains 15,844 pathways, distributed over the complexes that yielded a resolved pathway; each pathway is stored as 100 milestone frames. Pathway share, dominant pathway, effective pathway number, path length, exit angle, path smoothness and the convergence residual are computed as defined in the descriptor dictionary released with the dataset, which lists every field together with its unit and implementation.

## Binding-pocket angiography

Averaging the bias potential over the replica ensemble estimates the three-dimensional free-energy surface, $F(Rc) \propto -\langle V(Rc) \rangle$, following the BPA framework introduced with DD-13M[13]. Each BPA is stored as a four-dimensional volume (x, y, z, F) and rendered as a spatial probability map of ligand residence during escape. The same volume supplies the pocket-volume and escape-volume descriptors of the fifth dimension of the descriptor matrix.

## System-level record and descriptor matrix

For each modelled system we computed the five descriptor dimensions described in Results and displayed in Fig. 2, with the algorithms given in the descriptor dictionary released with the dataset. Descriptors were computed with a versioned, containerised workflow implemented with MDAnalysis[40], NumPy[41] and scikit-learn[42], stored with a provenance record and a completeness flag, and released with the trajectory. Missing values were retained as explicit nulls. Structural and dynamic descriptors use least-squares superposition[38] and Shrake–Rupley surface integration[39]. The bias work accumulated along each replica is stored as four summary statistics. These are the work values that enter the Jarzynski equality[43], but no exponential average and therefore no free-energy estimate is formed here: the trajectories are biased, the replica count is small relative to the variance of the work, and the values are exploratory.

## Group definition, construct-aware splitting and anomaly flags

Each structure was mapped to its UniProt accession[47], and structures sharing an accession were aggregated. This yielded 3,687 groups. Within each group, structures were aligned with TM-align[48]; systems in which the ligand centre of mass lay outside the protein surface at the first frame were excluded from pathway analysis; and groups containing distinct construct boundaries or oligomeric states were split into structural classes before pathway statistics were computed (1→N). Structures from different accessions that aligned closely were flagged as N-in-1 candidates and superposed for qualitative comparison. Each complex receives a system tag following the DD-13M scheme (manual setup, shallow pocket, short pathway, single pathway or multiple pathway). Each of the top-100 groups is described by a family record whose primary annotation is seven columns: the accession, its pocket class (open or pathway), its pathway multiplicity (single or multiple), the number of structures, its representativeness, its degeneracy flag and its domain-separation flag; quantitative supplements record the group size and rank, the degenerate partners and group identifier, the class sizes and number of domains, and the per-pathway composition. Degeneracy and domain separation are annotations only and do not alter the pathway statistics. We do not impose a higher-level family taxonomy because such groupings are not standardised for this purpose.

### Statistical analysis

The analyses reported here are descriptive and graphical. Pathway composition is reported as the replica fraction of each cluster, $p_i = n_i / \Sigma_j n_j$, and summarised by the dominant share $p_{max}$, the dominant-pathway identity and the number of effective pathways. Affinity trends are summarised by ordinary least-squares regression of log10(Kd/nM) on ligand long-axis length or ligand mass, with the coefficient of determination $R^2$ reported; no P values are attached to these fits because the residuals are heteroscedastic and the sampling is not random with respect to ligand chemistry. The separation of the P03366 structural classes in the affinity–size plane is summarised by the accuracy of a single ligand-length threshold, (TP + TN)/N, evaluated on the annotated class labels. Subgroup comparisons across dominant pathways are reported as distributions, not as significance tests, because the C2 and C3 subsets contain 7 and 4 ligands. Effective pathway counts use a fixed share cut-off, as specified in the released descriptor dictionary. Pathway shares are replica fractions under a fixed protocol, not equilibrium branching probabilities: a share of p carries a binomial standard error of $\sqrt{(p(1-p)/50)}$, which is 7 percentage points at p = 0.5. Per-ligand compositions are therefore read at the level of trends, and statements about pathway shares are made for groups rather than for single complexes.

## Data availability

Trajectories, modelled structures, pathways, BPA maps and the descriptor matrix (including work distributions) are archived in MDRepo[14] under the Dissociation Dynamic Database collection (publication record: https://mdrepo.org/pub/view/158); each complex has a stable accession (for example, MDR00105481). The complete set of 15,540 complex-level records is public. The project portal (https://aimm.szbl.ac.cn/database/ddd/#/home?version=DD03B) provides search, visualisation and bulk download. The resource is free, requires no registration, and will be maintained for at least five years from publication.

## Code availability

Analysis code and descriptor definitions are archived with the dataset in the MDRepo collection listed under Data availability (https://mdrepo.org/pub/view/158); the scripts used to reproduce the statistics in Figs. 3–6 are available from the corresponding author.

## References

1. Copeland, R. A., Pompliano, D. L. & Meek, T. D. Drug–target residence time and its implications for lead optimization. *Nat. Rev. Drug Discov.* **5**, 730–739 (2006).

2. Lu, H. & Tonge, P. J. Drug–target residence time: critical information for lead optimization. *Curr. Opin. Chem. Biol.* **14**, 467–474 (2010).

3. Wang, R., Fang, X., Lu, Y., Yang, C. Y. & Wang, S. The PDBbind database: methodologies and updates. *J. Med. Chem.* **48**, 4111–4119 (2005).

4. Liu, Z. et al. Forging the basis for developing protein–ligand interaction scoring functions. *Acc. Chem. Res.* **50**, 302–309 (2017).

5. Liu, Z. et al. PDB-wide collection of binding data: current status of the PDBbind database. *Bioinformatics* **31**, 405–412 (2015).

6. Meyer, T. et al. MoDEL: a database of atomistic molecular dynamics trajectories. *Structure* **18**, 1399–1409 (2010).

7. Rodríguez-Espigares, I. et al. GPCRmd uncovers the dynamics of the 3D-GPCRome. *Nat. Methods* **17**, 777–787 (2020).

8. Siebenmorgen, T. et al. MISATO: machine learning dataset of protein–ligand complexes for structure-based drug discovery. *Nat. Comput. Sci.* **4**, 367–378 (2024).

9. Vander Meersche, Y., Cretin, G., Gheeraert, A., Gelly, J. C. & Galochkina, T. ATLAS: protein flexibility description from atomistic molecular dynamics simulations. *Nucleic Acids Res.* **52**, D384–D392 (2024).

10. Mokhtari, O., Bignon, E., Khakzad, H. & Karami, Y. DynaRepo: the repository of macromolecular conformational dynamics. Preprint at *bioRxiv* https://doi.org/10.1101/2025.08.14.670260 (2025).

11. Korlepara, D. B. et al. PLAS-20k: extended dataset of protein–ligand affinities from MD simulations for machine learning applications. *Sci. Data* **11**, 180 (2024).

12. Charron, N. E. et al. Navigating protein landscapes with a machine-learned transferable coarse-grained model. *Nat. Chem.* **17**, 1284–1292 (2025).

13. Li, M. et al. A novel 4-D dataset paradigm for studying complete ligand–protein dissociation dynamics. Preprint at *arXiv* https://doi.org/10.48550/arXiv.2504.18367 (2025).

14. Roy, A. et al. MDRepo—an open data warehouse for community-contributed molecular dynamics simulations of proteins. *Nucleic Acids Res.* **53**, D477–D486 (2025).

15. Berman, H. M. et al. The Protein Data Bank. *Nucleic Acids Res.* **28**, 235–242 (2000).

16. Branduardi, D., Gervasio, F. L. & Parrinello, M. From A to B in free energy space. *J. Chem. Phys.* **126**, 054103 (2007).

17. Smith, Z. et al. Toward automated physics-based absolute drug residence time predictions. *J. Chem. Inf. Model.* **65**, 13360–13373 (2025).

18. Okimoto, N., Suenaga, A. & Taiji, M. Evaluation of protein–ligand affinity prediction using steered molecular dynamics simulations. *J. Biomol. Struct. Dyn.* **35**, 3221–3231 (2017).

19. Nguyen, H. L., Thai, N. Q. & Li, M. S. Determination of multidirectional pathways for ligand release from the receptor: a new approach based on differential evolution. *J. Chem. Theory Comput.* **18**, 3860–3872 (2022).

20. Patel, J. S., Berteotti, A., Ronsisvalle, S., Rocchia, W. & Cavalli, A. Steered molecular dynamics simulations for studying protein–ligand interaction in cyclin-dependent kinase 5. *J. Chem. Inf. Model.* **54**, 470–480 (2014).

21. Grubmüller, H., Heymann, B. & Tavan, P. Ligand binding: molecular mechanics calculation of the streptavidin–biotin rupture force. *Science* **271**, 997–999 (1996).

22. Conflitti, P. et al. Functional dynamics of G protein-coupled receptors reveal new pathways for drug discovery. *Nat. Rev. Drug Discov.* **24**, 251–275 (2025).

23. Limongelli, V., Bonomi, M. & Parrinello, M. Funnel metadynamics as accurate binding free-energy method. *Proc. Natl Acad. Sci. USA* **110**, 6358–6363 (2013).

24. Raniolo, S. & Limongelli, V. Ligand binding free-energy calculations with funnel metadynamics. *Nat. Protoc.* **15**, 2837–2866 (2020).

25. Huang, Y.-P. et al. SPONGE: a GPU-accelerated molecular dynamics package with enhanced sampling and AI-driven algorithms. *Chin. J. Chem.* **40**, 160–168 (2022).

26. Xia, Y. & Gao, X. Xponge: a Python package to perform pre- and post-processing of molecular simulations. *J. Open Source Softw.* **7**, 4467 (2022).

27. Maier, J. A. et al. ff14SB: improving the accuracy of protein side chain and backbone parameters from ff99SB. *J. Chem. Theory Comput.* **11**, 3696–3713 (2015).

28. Wang, J., Wolf, R. M., Caldwell, J. W., Kollman, P. A. & Case, D. A. Development and testing of a general amber force field. *J. Comput. Chem.* **25**, 1157–1174 (2004).

29. Jorgensen, W. L., Chandrasekhar, J., Madura, J. D., Impey, R. W. & Klein, M. L. Comparison of simple potential functions for simulating liquid water. *J. Chem. Phys.* **79**, 926–935 (1983).

30. Loncharich, R. J., Brooks, B. R. & Pastor, R. W. Langevin dynamics of peptides: the frictional dependence of isomerization rates of N-acetylalanyl-N′-methylamide. *Biopolymers* **32**, 523–535 (1992).

31. Andersen, H. C. Molecular dynamics simulations at constant pressure and/or temperature. *J. Chem. Phys.* **72**, 2384–2393 (1980).

32. Laio, A. & Parrinello, M. Escaping free-energy minima. *Proc. Natl Acad. Sci. USA* **99**, 12562–12566 (2002).

33. Wang, J., Ishchenko, A., Zhang, W., Razavi, A. & Langley, D. A highly accurate metadynamics-based dissociation free energy method to calculate protein–protein and protein–ligand binding potencies. *Sci. Rep.* **12**, 2024 (2022).

34. Valsson, O., Tiwary, P. & Parrinello, M. Enhancing important fluctuations: rare events and metadynamics from a conceptual viewpoint. *Annu. Rev. Phys. Chem.* **67**, 159–184 (2016).

35. Barducci, A., Bussi, G. & Parrinello, M. Well-tempered metadynamics: a smoothly converging and tunable free-energy method. *Phys. Rev. Lett.* **100**, 020603 (2008).

36. Ward, J. H. Hierarchical grouping to optimize an objective function. *J. Am. Stat. Assoc.* **58**, 236–244 (1963).

37. Henkelman, G., Uberuaga, B. P. & Jónsson, H. A climbing image nudged elastic band method for finding saddle points and minimum energy paths. *J. Chem. Phys.* **113**, 9901–9904 (2000).

38. Kabsch, W. A solution for the best rotation to relate two sets of vectors. *Acta Crystallogr. A* **32**, 922–923 (1976).

39. Shrake, A. & Rupley, J. A. Environment and exposure to solvent of protein atoms. Lysozyme and insulin. *J. Mol. Biol.* **79**, 351–371 (1973).

40. Gowers, R. J. et al. MDAnalysis: a Python package for the rapid analysis of molecular dynamics simulations. In *Proc. 15th Python in Science Conf.* 98–105 (2016).

41. Harris, C. R. et al. Array programming with NumPy. *Nature* **585**, 357–362 (2020).

42. Pedregosa, F. et al. Scikit-learn: machine learning in Python. *J. Mach. Learn. Res.* **12**, 2825–2830 (2011).

43. Jarzynski, C. Nonequilibrium equality for free energy differences. *Phys. Rev. Lett.* **78**, 2690–2693 (1997).

44. Sillitoe, I. et al. CATH: increased structural coverage of functional space. *Nucleic Acids Res.* **49**, D266–D273 (2021).

45. Andreeva, A., Kulesha, E., Gough, J. & Murzin, A. G. The SCOP database in 2020: expanded classification of representative family and superfamily domains of known protein structures. *Nucleic Acids Res.* **48**, D376–D382 (2020).

46. Schaeffer, R. D., Liao, Y., Cheng, H. & Grishin, N. V. ECOD: new developments in the evolutionary classification of domains. *Nucleic Acids Res.* **45**, D296–D302 (2017).

47. UniProt Consortium. UniProt: the universal protein knowledgebase in 2023. *Nucleic Acids Res.* **51**, D523–D531 (2023).

48. Zhang, Y. & Skolnick, J. TM-align: a protein structure alignment algorithm based on the TM-score. *Nucleic Acids Res.* **33**, 2302–2309 (2005).